\documentclass[trackchanges]{aastex7}

\usepackage{amsmath}
\usepackage{svg}
\usepackage{subcaption}
\usepackage{comment}

\begin{document}

\title{Thermodynamic Evolution of Flaring Loops with Non-local Thermal Transport}

\author[orcid=0000-0002-3505-9542
,gname=Sergey, sname='Belov']{S. Belov} 
\affiliation{Centre for Fusion, Space and Astrophysics, Department of Physics, University of Warwick,  Coventry CV4 7AL, UK}
\email{Sergey.Belov@warwick.ac.uk}

\author[orcid=0009-0008-2585-9280, gname=Thomas, sname='Parmenter']{T. Parmenter} 
\affiliation{Centre for Fusion, Space and Astrophysics, Department of Physics, University of Warwick,  Coventry CV4 7AL, UK}
\email{thomas.parmenter@warwick.ac.uk}

\author[orcid=0000-0002-9322-4913,
gname=Tony, sname='Arber']{T. Arber} 
\affiliation{Centre for Fusion, Space and Astrophysics, Department of Physics, University of Warwick,  Coventry CV4 7AL, UK}
\email{}

\author[gname=Dmitrii, sname='Kolotkov', orcid=0000-0002-0687-6172]{D. Kolotkov} 
\affiliation{Centre for Fusion, Space and Astrophysics, Department of Physics, University of Warwick,  Coventry CV4 7AL, UK}
\email{}
\affiliation{Engineering Research Institute \lq\lq Ventspils International Radio Astronomy Centre (VIRAC)\rq\rq, Ventspils University of Applied Sciences, Ventspils, LV-3601, Latvia}

\author[orcid=0000-0002-1820-4824,gname=Fabio, sname='Reale']{F. Reale} 
\affiliation{Dipartimento di Fisica e Chimica, Università di Palermo, Piazza del Parlamento 1, I-90134 Palermo, Italy}
\email{}

\author[orcid=0000-0003-0784-1294
,gname=Tom, sname='Goffrey']{T. Goffrey} 
\affiliation{Centre for Fusion, Space and Astrophysics, Department of Physics, University of Warwick,  Coventry CV4 7AL, UK}
\email{}

\collaboration{all}{}

\begin{abstract}

Hot solar coronal loops, such as flaring loops, reach temperatures where the thermal transport becomes non-local. This occurs when the mean-free-path of electrons can no longer be assumed to be small.
Using a modified version of the Lare2d code, we study the evolution of flare-heated coronal loops under three thermal transport models: classical Spitzer-H\"{a}rm (SH), a flux-limited local model (FL), and the non-local Schurtz–Nicolaï–Busquet (SNB) model. 
The SNB model is used extensively in laser-plasma studies. It has been benchmarked against accurate non-local Vlasov-Fokker-Planck models and proven to be the most accurate non-local model which can be applied on fluid time-scales.
Analysis of the density–temperature evolution cycles near the loop apex reveals a distinct evolutionary path for the SNB model, with higher temperatures and lower densities than local models. During energy deposition, the SNB model produces a more localised and intense temperature peak at the apex due to heat flux suppression, which also reduces chromospheric evaporation and results in lower post-flare densities. EUV emission synthesis shows that the SNB model yields flare light curves with lower peak amplitudes and smoother decay phases. We also find that non-local transport affects equilibrium loop conditions, producing hotter and more rarefied apexes. These findings emphasise the need to account for non-local conduction in dynamic solar phenomena and highlight the potential of the SNB model for improving the realism of flare simulations. Flux-limited conduction models cannot reproduce the results of non-local transport 
covered by the SNB model.

\end{abstract}

\keywords{}


\section{Introduction} 

Solar flares are the most intensively studied phenomena in heliophysics \citep[see, e.g., reviews][]{Shibata2011, Benz2016}. During flares, large amounts of energy are rapidly released into localised regions of the solar atmosphere, leading to plasma heated to higher temperatures. Under these conditions,  the thermal electron mean free path $\lambda_T$ can become comparable to the characteristic temperature length scale $L_T = T / |\nabla T|$ of the plasma. In this case, the classical model of local thermal transport \citep{Spitzer1953} does not correctly describe heat transport, and non-local thermal transport effects should be taken into account. These effects are preheating when the effective spatial profile of heat flux is broadened by hotter electrons carrying energy for longer distances, and suppression of the peak heat
flux. Studies show that the classical thermal transport description starts to fail when the Knudsen number $K_n = \lambda_T / L_T > 0.01$ \citep{Arber2023} or even $K_n > 0.001$ \citep{Gray1980, Belov2025}. \cite{Battaglia2009} demonstrated that the local thermal transport is violated under typical solar flare values (see Fig. 4).

There are few studies modelling thermal transport as non-local despite the evidence that the local thermal transport is violated under coronal and, especially, flaring conditions. \cite{Karpen1987, Karpen1989}  modelled a 1D flaring solar atmosphere, comparing classical, flux-limited, and non-local transport models. The latter produced significantly different temperature, density, and velocity profiles during the rise phase of the flare compared to local models. In particular, non-local thermal transport led to significantly higher coronal temperature, and in earlier, but weaker, chromospheric evaporation. \cite{Karpen1987} concluded that non-local thermal transport could strongly affect the coronal, transition region, and chromospheric parameters during flares. Later, \cite{ciaravella1991non} studied the effect of non-local transport in static coronal loops and showed that its influence on the distribution of differential emission measure was important for loops longer than the
pressure scale height. \cite{West2004} demonstrated that the coronal cooling times due to thermal conduction are increased when the non-local thermal transport is taken into account. \cite{Silva2018} performed a 3D magnetohydrodynamic (MHD) simulation of the active region for the local and non-local thermal transport models. A smoother temperature profile was obtained for the non-local model, which also resulted in two times higher temperatures in some of the regions in the lower corona.

To describe non-local thermal transport, all these studies utilised the kernel model derived for laser plasmas \citep{Luciani1983} commonly known as the LMV model. In this model, the non-local heat flux is obtained by convolving the local heat flux with a non-local kernel. In recent years, the model first derived by \cite{Schurtz2000} (SNB) has emerged as the most effective way to treat non-local thermal transport in MHD simulations of laser plasmas. It has been shown to produce a heat flux with good agreement to results obtained from simulations using the full Vlasov-Fokker-Planck (VFP) equation \citep{Marocchino2013, Brodrick2017, Sherlock2017}. While the SNB model does not capture the full detail of a VFP solution, which remains computationally impractical for 3D simulations, it reproduces the non-local preheat and flux-limiting properties well for temperature profiles typical of the solar corona. The SNB model requires the solution of PDEs and is more straightforward to implement numerically than the LMV approach, which requires tracing magnetic field lines that, in the case of 3D simulations, leads to a sufficient number of these lines to reconstruct the temperature field. Moreover, SNB is a self-contained predictive theory that avoids the need to choose kernel functions or to set free parameters, unlike the LMV model. However, care must still be taken during extremely large flares, as $K_n$ can become much larger than 1, and the SNB applicability is violated. In such cases, alternative models of non-local energy transport  must be considered. 

In this Letter, we present for the first time solutions for flaring loops using the SNB model. Based on these results we propose the SNB model as an efficient and accurate framework for modelling non-local thermal energy transport. Importantly this is restricted to thermal transport and thus requires that the electron distribution is not far from Maxwellian allowing the definition of a thermal temperature. For this reason SNB is a good model for MHD studies of the solar corona where only fluid temperatures are known.

\section{Model} \label{sec:model}
In this study, we consider {a 1D model of a semicircular loop} with a total length in
the corona of $50$\,Mm.
{The cross-section is assumed to be non-varying and to have no role in the considered model.}
{To provide enough material for chromospheric evaporation and mitigate the boundary effects \citep{Reale2014}}, the model atmosphere is vertically extended an additional $10$ Mm at chromospheric temperatures and densities below each loop
coronal footpoint
\citep[see, e.g.,][]{Klimchuk1987, Kucera2024}. To model plasma dynamics along the loop, we numerically solve the equations of mass, momentum, and energy transport using the Lare2d code \citep{Arber2001}:

\begin{equation}
\label{eq:mass_tr}
\frac{\mathrm{\partial} \rho}{\mathrm{\partial} t} + v\frac{\mathrm{\partial} \rho}{\mathrm{\partial} s} = - \rho \frac{\mathrm{\partial} v}{\mathrm{\partial} s},
\end{equation}
\begin{equation}
\label{eq:momentum_tr}
\frac{\mathrm{\partial} v}{\mathrm{\partial} t} + v\frac{\mathrm{\partial} v}{\mathrm{\partial} s} = - \frac{1}{\rho} \frac{\mathrm{\partial} P}{\mathrm{\partial} s} - g_{||},
\end{equation}
\begin{equation}
\label{eq:energy_tr}
\rho \frac{\mathrm{\partial} \epsilon}{\mathrm{\partial}t} + \rho v\frac{\mathrm{\partial} \epsilon}{\mathrm{\partial} s} = -P \frac{\mathrm{\partial} v}{\mathrm{\partial} s} - \frac{\mathrm{\partial} q}{\mathrm{\partial}s} - n_e^2\Lambda(T) + H.
\end{equation}
In Equations (\ref{eq:mass_tr})--(\ref{eq:energy_tr}), $s$ is the 1D spatial coordinate along the loop, $v$ is the parallel velocity, $P$ is the gas pressure, {$g_{||}=g\cos\left(\pi\left(s-s_{\mathrm{legs}}\right)/\left(s_{\mathrm{max}}-2s_{\mathrm{legs}}\right)\right)$ is the parallel gravitational acceleration in the semicircular part of the loop, where $g=274$\, m/s$^2$ is the surface gravitational acceleration, $s_{\mathrm{legs}}=10$\,Mm is the length of each chromospheric leg, $s_{\mathrm{max}}=70$\,Mm is the domain size. In the chromospheric regions ($s < s_{\mathrm{legs}}$ and $s > s_{\mathrm{max}} - s_{\mathrm{legs}}$), the parallel gravitational acceleration is taken as constant: $g_{||} = g$ on the left leg and $g_{||} = -g$ on the right leg. Here,} $\epsilon = P / \rho(\gamma - 1)$ is the specific internal energy density, $q$ is the heat flux, $n_e$ is the electron number density, $\Lambda(T)$ is the radiative loss function for an optically thin plasma obtained using the CHIANTI database \citep{Dere1997, DelZanna2021}, and $H$ is a heating term. A fully ionised plasma with mean ion mass of $1.2m_p$, where $m_p$ is the proton mass, was assumed.

The heating term $H$ consists of two parts: a constant uniform heating $H_0$ supporting the steady-state of the loop, and a heating term $H_1\left(t, s\right)=Q_{\mathrm{max}}\,A\left(t\right)\exp\left(-\left(s-s_0\right)^2/\sigma^2\right)$ representing energy release during a flare. Here, we choose $H_0=4.0\times10^{-4}$\,W/m$^3$ to have the apex temperature around $3$ MK \citep{Rosner1978}. For the flare heating term $H_1\left(t, s\right)$, we use a triangular symmetric time profile $A\left(t\right)$ with duration $t_{\mathrm{flare}}=600$\,seconds, and choose the spatial flare size $\sigma=5$\,Mm and a flare location $s_0=35$\,Mm corresponding to a flare at the loop apex (the initial coronal loop extends from $s=10$\,Mm to $s=60$\,Mm). We set the flare amplitude to
$Q_{\mathrm{max}}=0.012$\,W/m$^3$, resulting in a total energy released during the flare of $Q_{\mathrm{flare}} = 0.5 \pi^{1/2}Q_{\mathrm{max}} \sigma   t_{\mathrm{flare}}\approx3.2\times 10^7$\,J/m$^2$. {{For this heating release and for the local thermal transport, the peak temperature can be estimated using the RTV scaling law \citep{Rosner1978} as $T_{\mathrm{peak}}\approx\left(10^{11}H_0 \lambda_0^2\right)^{2/7}=6.37$\,MK, where $H_0=Q_{\mathrm{flare}}/2\sigma t_{\mathrm{flare}}$, and $\lambda_0=0.5s_{\mathrm{max}}$}}, {{which then represents a relatively small flare.}}

We conduct our numerical study for three different models of heat flux $q$: the local thermal transport model $q_{\mathrm{\mathrm{SH}}}$ described by the Spitzer-H\"{a}rm (SH) approximation \citep{Spitzer1953} valid for shorter electron mean-free path; the flux-limited local thermal transport model (FL) $q_{\mathrm{\mathrm{FL}}}$ \citep{OLSON2000619} limiting a maximum value of heat flux as $\alpha q_{\mathrm{fs}}$, where $\alpha$ is a free parameter, and $q_{\mathrm{fs}}$ is a free-streaming heat flux; and the Schurtz-Nicola{\"i} -Busquet (SNB) model \citep{Schurtz2000, Brodrick2017} $q_{\mathrm{\mathrm{SNB}}}$ describing nonlocal thermal transport {{and representing a  solution of the VFP equation approximated in energy bins}} \citep[see][for the brief comparison between the SH, FL and SNB thermal transport models]{Arber2023, Belov2025}. In this study, the flux-limiter $\alpha$ is set to 0.06 for the FL model. For the SNB model, 100 logarithmically spaced temperature groups were used in the interval between 0 and $20T_{\mathrm{max}}$, where $T_{\mathrm{max}}$ is the maximum temperature of the domain on each time step.

In our simulations, we use 1400 uniformly spaced grid points along the loop axis. Zero velocity boundary conditions are imposed, whilst the temperature and density are held constant at their initial values. As the extended chromospheric legs at 10\,Mm
deep no velocity perturbation or heat flux from the flare release reach these boundaries. We initialise our simulations with a parabolic profile:

\begin{equation}
\label{eq:in_state}
T(s) =
\begin{cases}
    as^2 +bs + c, &  s_{\mathrm{legs}} \leq s \leq s_{\mathrm{max}}-s_{\mathrm{legs}}, \\
    T_{\mathrm{ch}}, & \text{otherwise}
\end{cases}
\end{equation}
\begin{equation}
a = -\frac{T_{\mathrm{cor}} - T_{\mathrm{ch}}}{(s_{\mathrm{legs}} - 0.5s_{\mathrm{max}})^2}, \quad
b = -a s_{\mathrm{max}}, \quad
c = T_{\mathrm{cor}} + 0.25a s_{\mathrm{max}}^2,
\end{equation}
where $T_{\mathrm{cor}}=2.2$\,MK is the initial coronal temperature, $T_{\mathrm{ch}}=20000$\,K is the temperature of the chromospheric legs. Given the initial temperature profile, Equation (\ref{eq:in_state}) {and assuming hydrostatic equilibrium ($\partial P/\partial s= -\rho g_{||}$, where $P=k_B \rho T / 0.6 m_p$), the initial density profile was calculated numerically using a second-order finite difference scheme.  These profiles satisfy hydrostatic balance with gravity but do not represent a full thermodynamic equilibrium, which would require a balance between thermal conduction, radiative losses, and background heating. To satisfy both hydrostatic and thermodynamic equilibria, we allowed the atmosphere to evolve numerically from this approximate hydrostatic state by solving the full time-dependent system, Equations (\ref{eq:mass_tr})–(\ref{eq:energy_tr}), with the flare heating term $H_1\left(t, s\right)$ switched off. The Lare2d code was run for 15,000 seconds until a steady state was reached.  We use this approach of reaching a steady state from the guessed state because an analytical form of equilibrium for the model considered is unavailable, and it cannot be obtained numerically by integrating the RHS of Equations (\ref{eq:mass_tr})–(\ref{eq:energy_tr}) due to the implicit nature of the SNB thermal transport model. Indeed, in the SNB model, the heat flux is inherently non-local, i.e. is derived individually by solving a spatial-dependent ODE (15) from \cite{Arber2023} for each temperature bin, so it cannot be expressed as a local function of temperature.  Thus, the used method ensures consistency with the discretised numerical scheme and provides a unified and robust treatment across different thermal transport models.}

\begin{figure*}[ht!]
\centering
\plotone{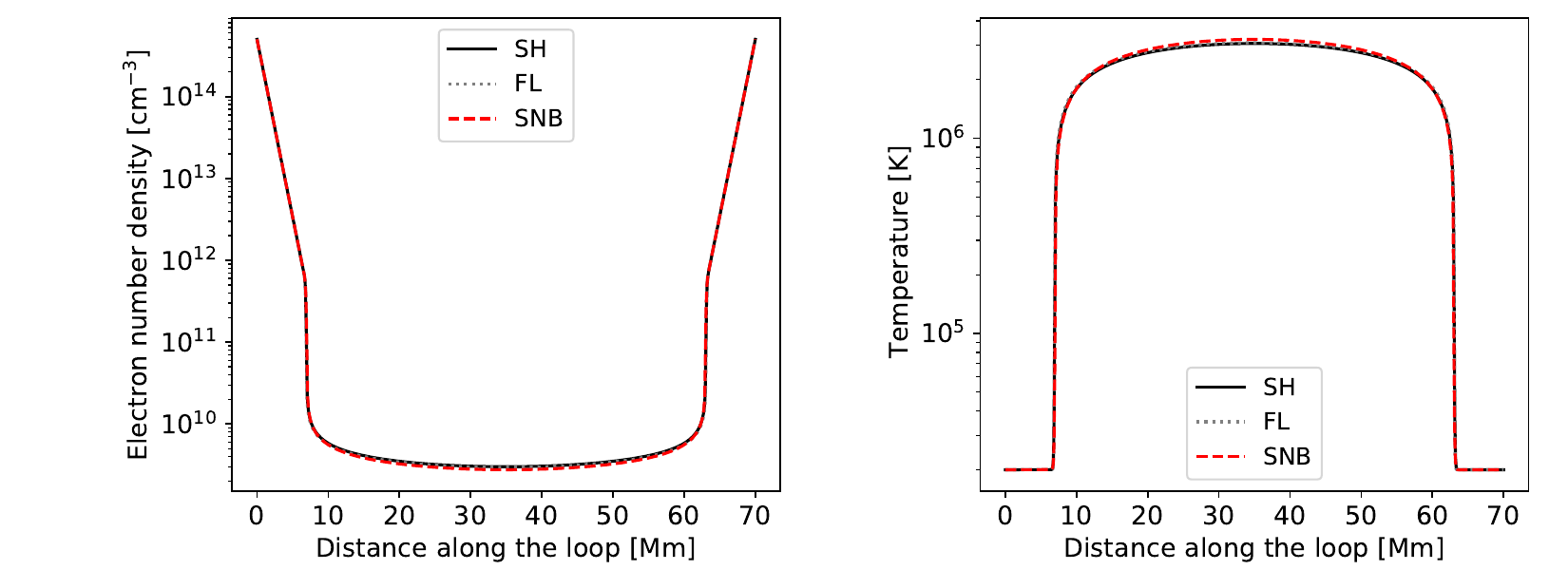}
\caption{Initial atmosphere before the flare for SH (black line), FL (grey dotted line), and SNB (red dashed line) thermal transport models. Note both vertical axes use logarithmic scales.}
\label{fig:in_state}
\end{figure*}

\begin{figure}
\centering
\includegraphics[width = 110mm]{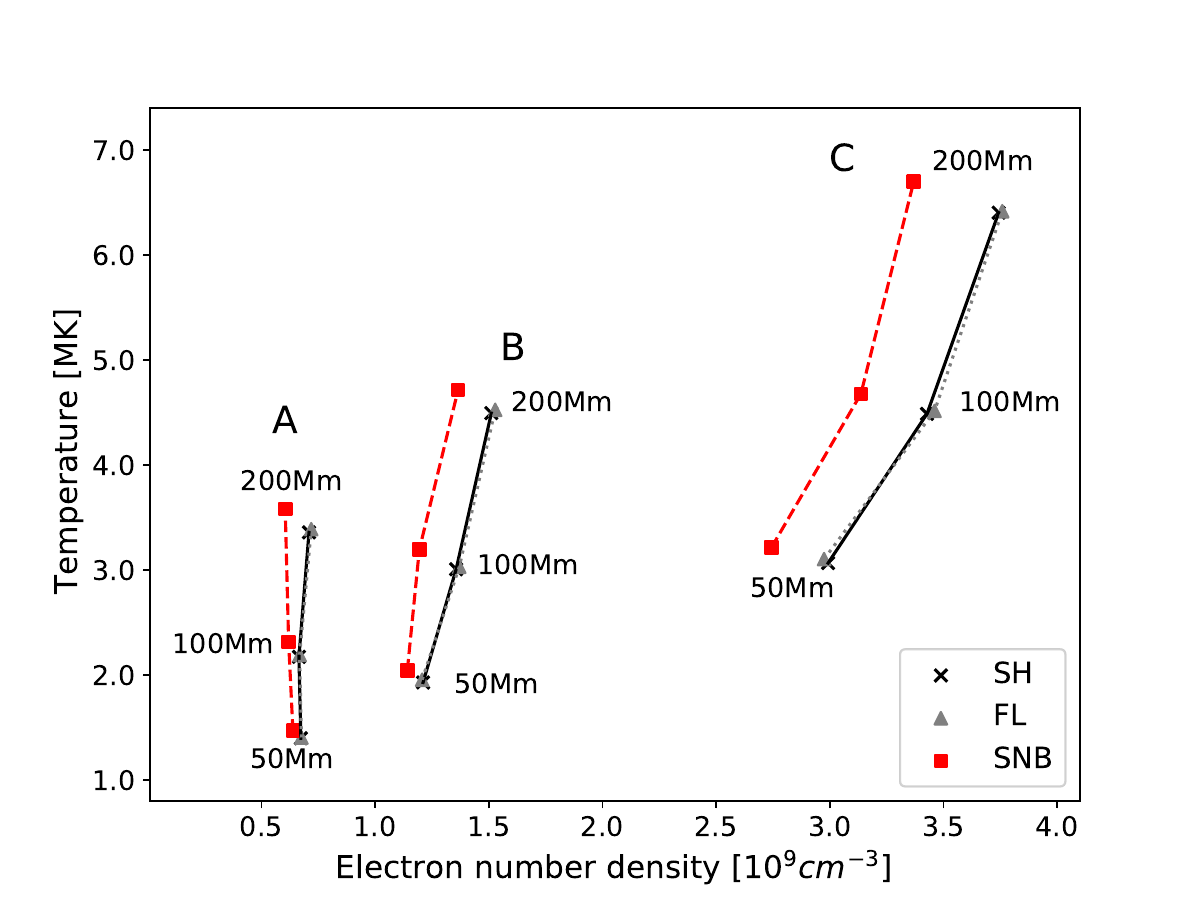}
\caption{Temperature and electron number density at the apex of the loop  for the SH (black crosses), FL (grey triangles), and SNB (red squares) thermal transport models, across a range of background heating rates $H_0$ and loop lengths $L$. Each group of connected points (A, B, and C) represents a family of solutions corresponding to a fixed heating rate $H_0$: $4.0 \times 10^{-5}$ \, Wm$^{-3}$ (A), $1.1 \times 10^{-4}$ \, Wm$^{-3}$ (B), and $4.2 \times 10^{-4}$ \, Wm$^{-3}$ (C). Within each group, lines connect points for loop lengths of 50\,Mm, 100\,Mm, and 200\,Mm for a given thermal transport model. The plot shows that SNB results in a lower density and higher temperature than SH (or FL), with this difference becoming more pronounced as $H_0$ is increased.
}
\label{fig:scaling}
\end{figure}

Fig.~\ref{fig:in_state} shows the steady-state distributions of density and temperature along the loop axis {satisfying both hydrostatic and thermodynamic equilibria}. It can be seen from this figure that all the thermal conduction mechanisms result in similar density and temperature distributions with temperature $\approx3$\,MK and electron number density $\approx3\times10^9$\,cm$^{-3}$ at the apex. However, the exact temperature and density scaling with loop length is different for each heating mechanism, providing hotter and more rarefied apexes for the non-local thermal transport case (see Fig.~\ref{fig:scaling}).

After obtaining the {steady atmosphere}, we switch on the flare heating $H_1\left(t, s\right)$ and run the simulations for $10000$ seconds. We include edge-based shock viscosity \citep{caramana1998formulations}, with $\nu_1=0.1$ and $\nu_2=0.1$, and the Transition Region Adaptive Conduction (TRAC) method \citep{Johnston2019, Johnston2021} to remove the influence of numerical resolution on the transition region which evolves in steady state.

\section{Results}\label{sec:results}

\begin{figure*}[ht!]
\plotone{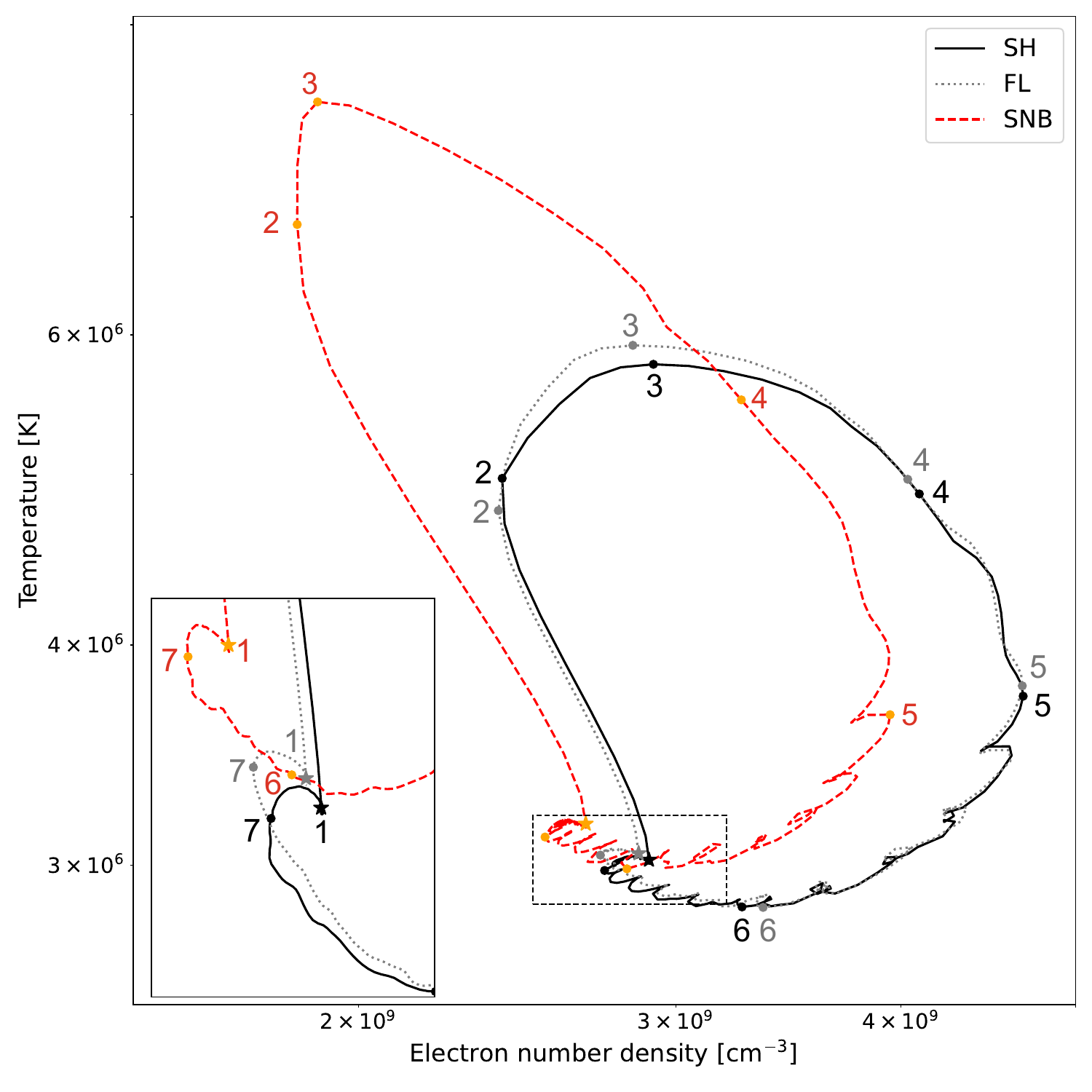}
\caption{The density-temperature thermodynamic cycles at the loop apex obtained for SH, FL, and SNB thermal transport models. Each cycle represents a plasma response to a heating event located at the loop apex and depositing $\approx3.2\times 10^7$\,J/m$^2$. The start of energy deposition is marked by star markers. The zoomed area has been smoothed over 9 points window (200 seconds) to increase readability.
\label{fig:cycles}}
\end{figure*}

After obtaining the {steady atmosphere}, we turn on the flare heating $H_1\left(t, s\right)$ and run simulations for each thermal transport mechanism for $10000$\,seconds. Fig. \ref{fig:cycles} shows density-temperature thermodynamic cycles obtaining by averaging the density and temperature inside upper $10$\,Mm of the loop (between $30$\,Mm and $40$\,Mm). The star markers denote the start of a heating deposition process. Each density-temperature cycle can be divided into seven parts, as indicated by numerical markers 1-7:
\begin{enumerate}
  \item Stage 1-2: The flare event starts, and thermal energy is deposited into the plasma. This leads to an increase in temperature and a decrease in density. At this stage, the trajectory corresponding to the non-local thermal transport model (SNB) deviates significantly from those of the SH and FL local thermal transport models. In the SNB case, the plasma is heated to higher temperatures, and, as a result, reaches lower densities. This can be explained by the fact that SNB suppresses heat fluxes causing more thermal energy to remain at the flare site.
  \item Stage 2-3: The flare continues heating the plasma, and the temperature reaches its peak value. However, the density starts increasing due to mass evaporation from the chromosphere caused by heat fluxes generated during the flare event. At this stage, the density increase is small for the SNB model because less thermal energy was transported to the chromosphere to drive evaporation during stage 1-2. 
  \item Stage 3-4: Plasma density continues to increase due to ongoing evaporation, while temperature decreases due to the combined effects of thermal conduction and growing radiative losses. For the local thermal transport models, the temperature does not change dramatically because the deposited thermal energy is quickly smoothed  by thermal conduction (which operates near an isothermal regime) and by increasing radiative cooling. In contrast, for the SNB case, the temperature drops more noticeably due to thermal conduction. This is  because the interplay between flare energy input and suppressed thermal conduction produces a more localised temperature peak. As a result, enhanced local temperature gradients lead to increased thermal conduction losses relative to the smoother temperature profiles seen in the local models. 
  \item Stage 4-5: The heat pulse ends, and the plasma continues cooling due to thermal conduction and radiative losses. At the same time, the plasma density increases further due to evaporation reaching its highest value at point 5. This value is higher for the local thermal transport mechanisms for the same reason as at point 2. Also, the generation of acoustic waves takes place at this stage, which is out of the scope of this paper \citep[see, e.g.,][]{Reale2019}.
  \item Stage 5-6: After reaching the maximum density, the thermal pressure is insufficient to support material in the corona against gravity, and the plasma begins to fall.
  \item Stage 6-7: At point 6, the temperature reaches its minimum and then starts increasing due to the dominance of background heating $H_0$ over conductive and radiative losses. 
  \item Stage 7-1: The loop returns to its initial state. During this process, the temperature remains almost constant, while the density increases due to chromospheric evaporation.
\end{enumerate}
Thus, comparing the density-temperature cycles for local and non-local thermal transport models, we find that the SNB model leads to markedly different density–temperature evolution compared to the SH and FL models. The key differences are a higher temperature during the flare stage, due to suppressed heat fluxes trapping more thermal energy in the corona, and a lower density resulting from stronger plasma expansion during the energy deposition stage and a reduced chromospheric evaporation rate.

\begin{figure*}[ht!]
\plotone{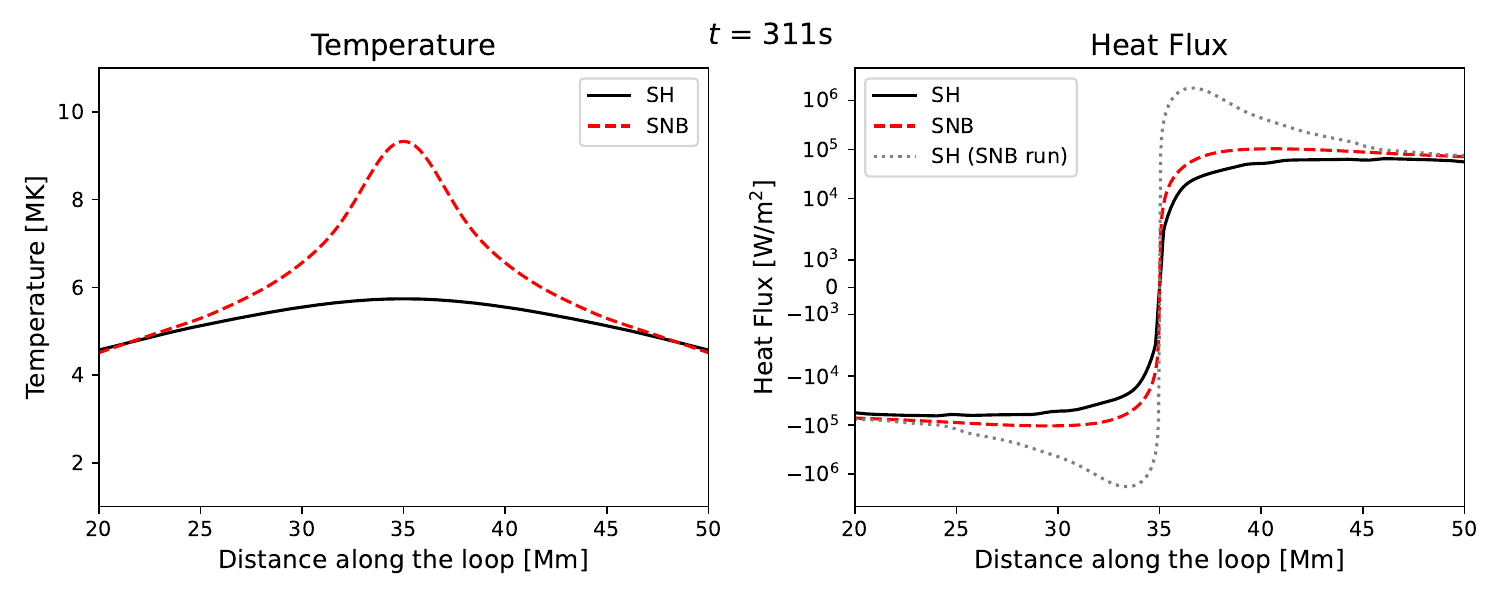}
\caption{Left panel: spatial profile of temperature near the flare peak (311 seconds after the start of heating) for SH and SNB thermal transport models. Right panel: corresponding spatial profiles of heat fluxes near the flare peak for the SH  (black solid line) and SNB (red dashed line). The grey dotted line shows the SH heat flux that would be present for the temperature and density profiles obtained from the SNB simulation. 
\label{fig:fluxes}}
\end{figure*}

Fig. \ref{fig:fluxes} shows the distribution of temperature (left panel) and heat flux (right panel) in the coronal part of the loop near the flare peak for the SH and SNB runs; in the right panel, the grey dotted line demonstrates what the SH heat flux could be for the temperature and density values obtained from the SNB run. Note that the heat fluxes were calculated using the open-access script\footnote{\url{https://github.com/Warwick-Solar/SNB_flux}} from \citep{Belov2025} applied to the output data. During the flare, the SNB model results in a localised temperature peak at the loop apex, in contrast, the temperature is more uniform in the SH case. As can be seen from the right panel of Fig. \ref{fig:fluxes}, the SNB heat flux is higher than the SH heat flux due to the presence of the localised temperature peak. However, this heat flux is still significantly smaller than it could be if using the SH model with the  temperature and density profiles obtained from the SNB run (compare the red dashed and grey dotted lines). 

\begin{figure*}[ht!]
\plotone{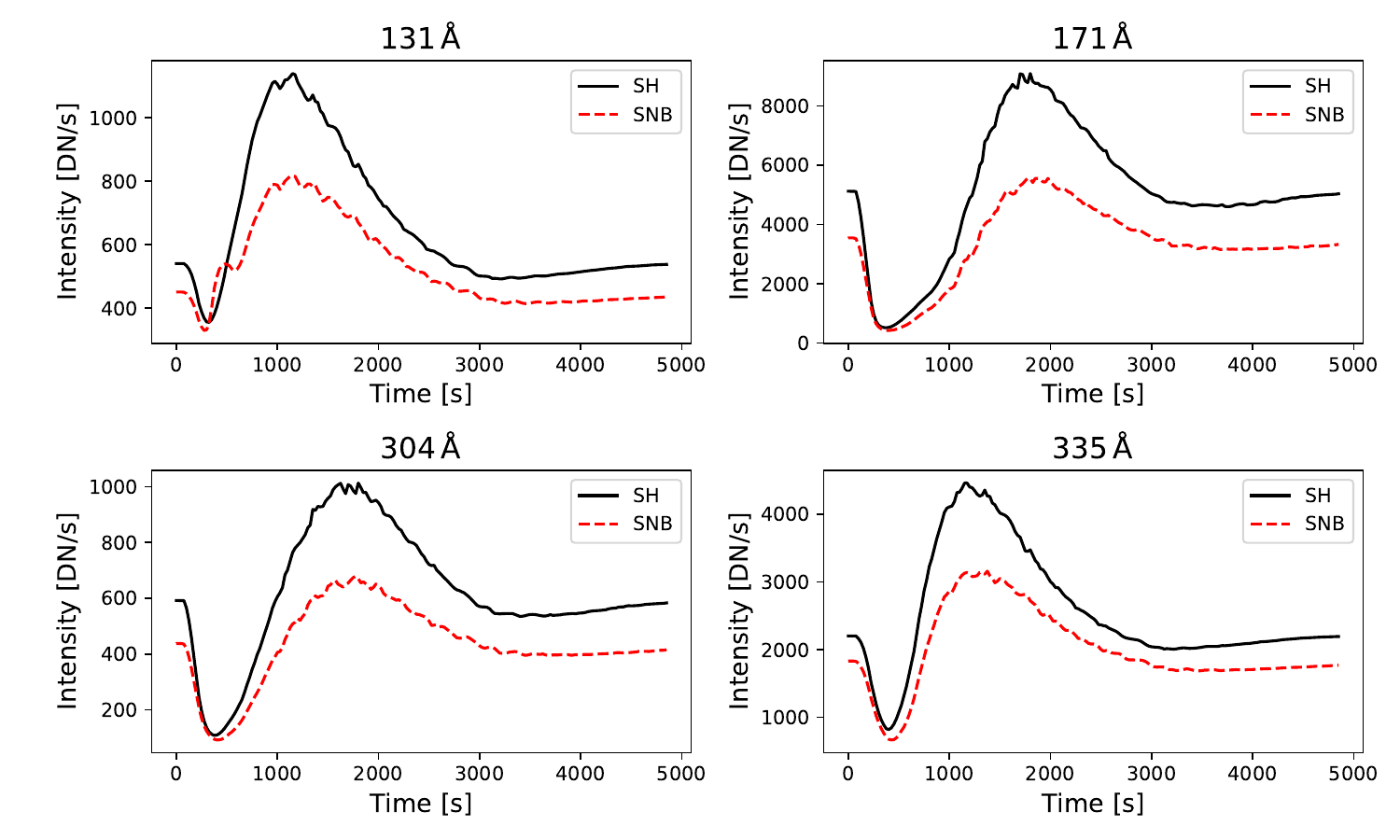}
\caption{The synthetic emission intensity summed up over the coronal part of the loop (15-55Mm) for four different EUV wavelengths (131\r{A}, 171\r{A}, 304\r{A}, 335\r{A}) generated using the SDO AIA response functions. 
\label{fig:l_curves}}
\end{figure*}

Fig. \ref{fig:cycles} demonstrates that the SNB thermal transport model results in a density-temperature cycle that significantly deviates from the SH case. This fact may lead to a change in extreme ultraviolet (EUV) emission which are sensitive to the density and temperature of the loop. Fig. \ref{fig:l_curves} shows the EUV emission integrated over a coronal part of the loop for the SH and SNB thermal transport models. The EUV emission intensity was estimated for 131\r{A}, 171\r{A}, 304\r{A}, and 335\r{A} {{channels}} as $n_e^2 G_{l}\left(T\right)\times2R_{\mathrm{loop}}$, where $G_{l}\left(T\right)$ is the SDO AIA response curve for a {{channel}} $l$, calculated using the FoMo reference tables \citep{VanDoorsselaere2016}, with { $R_{\mathrm{loop}}=1$\,Mm taken as an example value.} All four {{channels}} show a decrease in emission intensity initially followed by a large increase as the loop density increases, which can be attributed to the position of the energy deposition at the loop apex. From this peak, the intensity gradually returns towards its equilibrium value. For all the {{channels}} considered, the SNB light curves have lower peaks due to the loop's lower density during its thermodynamic evolution. The 131\r{A} curve exhibits an additional smaller maximum point near the start of the flare when using SNB rather than a smooth increase as seen with SH. This could be due to the higher maximum temperature predicted by the SNB model reaching a maximum of 131\r{A} response curve. In addition, the SNB light curves exhibit more shallow slopes for both the rise and decay phases compared to the SH case \citep[cf.][]{Kashapova2021, Kuznetsov2021}. During the rise phase, the increase in the EUV emission is supported by the chromospheric evaporation driven by thermal energy transport from a flare site. However, the suppressed heat fluxes in the SNB case lead to a slower evaporation, and so a more gradual rise. In the decay phase, the more shallow slope in the light curve for the SNB model might be attributed to heat flux limiting from non-local thermal transport, which increases the time for density and temperature to reduce following the flare. It is also worth noting that due to the lower equilibrium density, the emission is slightly reduced when using SNB, even before the flare.

\section{Discussion} 

In this study, a modified Lare2d code has been used to model a coronal loop's response to solar flare heating events. These simulations were carried out with three different models of thermal conduction: the classical Spitzer-H\"{a}rm theory (SH) \citep{Spitzer1953}, the local heat flux that is limited to some fraction $\alpha$ of the free-streaming flux (FL) \citep{OLSON2000619}, and the non-local model first presented by Schurtz, Nicolai and Busquet (SNB) \citep{Schurtz2000}. Before simulating flare energy release, we briefly considered how the thermal transport models affect the scaling of a coronal loop's apex temperature and density when in an equilibrium state. SNB results in a slightly hotter and more rarefied loop, with the difference between the models becoming larger as the coronal heating term is increased (see Fig. \ref{fig:scaling}).

For each flare simulation, we tracked the density-temperature evolution around the loop apex as shown in Fig. \ref{fig:cycles}. These thermodynamic cycles reveal that the non-local model results in an evolutionary path with higher temperatures and lower densities. During the flare energy deposition stage heat flux limiting leads to a higher and more localised temperature perturbation (see also the left panel in Fig. \ref{fig:fluxes}) compared to the local models considered. The lower density values can be explained by the stronger plasma expansion during the energy deposition stage and reduced chromospheric evaporation rate.
{{A future parametric study of the revealed difference between the SNB and local models upon the injected flare energy seems of great interest.}}
{Observationally, these cycles can be revealed, for example, with the use of imaging EUV instruments allowing for the Differential Emission Measure (DEM) reconstruction.}

Our results indicate that the simple FL approach cannot replicate the SNB results especially with respect to the density found within the loop. An alternative value of the FL could give closer agreement to SNB, however, this is a key limitation of the FL model. There is no single value for $\alpha$ suitable for use in all cases and the FL approach has limited predictive value, since $\alpha$ must be tuned to fit observations or more accurate simulations once the answer is known. 

A difference in EUV emission was found when using SNB compared to SH, with a lower intensity found throughout most of the flare's evolution as shown in Fig. \ref{fig:l_curves}. The shape of the emission curves is largely the same between the two models, with the maximum and minimum intensities occurring at approximately the same time. However, for 131\r{A} SNB gives an additional smaller maximum, which SH does not predict, potentially due to the larger peak temperature during simulations with the SNB model.  At shorter wavelengths, such as for soft x-rays, this difference could become more pronounced and so would be interesting to investigate further in future work. Additionally, the more localised temperature profile in the SNB case may be an observational evidence of non-local thermal transport effects in spatially resolved observations. {If spectral EUV instruments are considered, non-local thermal transport might manifest itself through the change in EUV line profiles, e.g., change in line widths and shifts. The study of this effect would require a dedicated work and may constitute an interesting follow-up of our work.}

With this Letter, we aim to highlight the importance of non-local thermal transport in the solar corona. We are motivated by the fact that the high temperatures and low densities found within the Sun's corona can cause the mean-free path to become very large, resulting in insufficient collisions for classical thermal conduction to apply. Therefore, accurately modeling heat transport within the Sun's atmosphere remains a challenge, especially during dynamic phenomena such as solar flares where large temperature gradients are present. Despite the fact that the first studies of non-local transport in coronal loops took place around forty years ago \citep{Karpen1987,Karpen1989, ciaravella1991non}, it has not been considered widely, partly because of the lack of accessible numerical tools such as SNB.  

\begin{acknowledgments}
  The work is funded by STFC Grant ST/X000915/1. DYK also acknowledges funding from the Latvian Council of Science Project No. lzp-2024/1-0023.
\end{acknowledgments}

\software{Lare2d \citep{Arber2001} 
}





\bibliography{refs}{}
\bibliographystyle{aasjournalv7}



\end{document}